 \documentclass[sigconf, authorversion]{acmart}

\copyrightyear{2023}
\acmYear{2023}
\setcopyright{acmlicensed}\acmConference[CHI '23]{Proceedings of the 2023 CHI Conference on Human Factors in Computing Systems}{April 23--28, 2023}{Hamburg, Germany}
\acmBooktitle{Proceedings of the 2023 CHI Conference on Human Factors in Computing Systems (CHI '23), April 23--28, 2023, Hamburg, Germany}
\acmPrice{15.00}
\acmDOI{10.1145/3544548.3580739}
\acmISBN{978-1-4503-9421-5/23/04}

\usepackage{colortbl}                  
\usepackage{framed}
\usepackage{soul}

\newcommand{\rframe}{($RQ_{\text{frm}}$)}
\newcommand{\rvis}{($RQ_{\text{vis}}$)}
\newcommand{\rint}{($RQ_{\text{int}}$)}

\newcommand{\update}[2]{#2}

\usepackage{enumitem}

\usepackage{blindtext}

\usepackage{wrapfig}

\begin{document}




\title{Designing Resource Allocation Tools to Promote Fair Allocation: \\ Do Visualization and Information Framing Matter?
}

\author{Arnav Verma}
\email{arnav@cs.toronto.edu}
\orcid{0000-0002-4018-9296}
\affiliation{
  \institution{University of Toronto}
  \city{Toronto}
  \country{Canada}
}

\author{Luiz Morais}
\email{luiz.morais@inria.fr}
\affiliation{
  \institution{Inria}
  \city{Bordeaux}
  \country{France}
}

\author{Pierre Dragicevic}
\email{pierre.dragicevic@inria.fr}
\affiliation{
    \institution{Université de Bordeaux, CNRS, Inria, LaBRI}
  \city{Bordeaux}
  \country{France}
}

\author{Fanny Chevalier}
\email{fanny@cs.toronto.edu}
\affiliation{%
 \institution{University of Toronto}
 \streetaddress{..}
 \city{Toronto}
 \country{Canada}
}

\renewcommand{\shortauthors}{Verma et al.}

\begin{abstract}
Studies on human decision-making focused on humanitarian aid have found that cognitive biases can hinder the fair allocation of resources. However, few HCI and Information Visualization studies have explored ways to overcome those cognitive biases. This work investigates whether the design of interactive resource allocation tools can help to promote allocation fairness. We specifically study the effect of presentation format (using text or visualization) and \update{information framing}{a specific framing strategy} (showing resources allocated to groups or individuals). In our three crowdsourced experiments, we provided different tool designs to split money between two fictional programs that benefit two distinct communities. Our main finding indicates that individual-framed visualizations and text may be able to curb unfair allocations caused by group-framed designs. This work opens new perspectives that can motivate research on how interactive tools and visualizations can be engineered to combat cognitive biases that lead to inequitable decisions.
\end{abstract}




\maketitle
\section{Introduction} \label{sec:intro}

\update{People tend to favor humanitarian causes to which they feel socially connected to ~\cite{caviola2021psychology}, leading to isolated communities in need of urgent care with fewer resources.}{}When a tornado tears down a town in Florida, North American donors and charities may provide an excess of monetary support to help with the rebuilding process while leaving little financial aid to Haiti, devastated by a similar but larger-scale disaster. \update{}{This situation can occur because people tend to be biased in favor of humanitarian causes which they feel socially connected to~\cite{caviola2021psychology}. This is problematic because some communities in need of urgent care might end up with less support than they should receive. This is an example of how \emph{cognitive biases} can negatively affect decision making: if decision makers only trust their intuition, available resources may end up poorly distributed.}



\update{}{
There are tools available to help or motivate people to donate. Donation websites such as \url{donorschoose.org} or fundraising platforms like \url{gofundme.com} act as tools which let users search for similar causes to which allocate their resources [5, 6], but comparison across these causes is not supported. Prior research has focused on encouraging people to contribute more towards a charity using tools such as conversational agents \cite{tran2022ask}, web interfaces \cite{mota2020desiderata, kuchler2020you}, and visual media like photographs and visualizations to elicit affective emotions such as compassion \cite{jenni1997explaining, dragicevic2022towards}. These aim to elicit \emph{more} donations, but little is known about tools to help people allocate available charitable resources more equally.
}

While there are many aspects to consider when making a resource allocation (such as the relative effectiveness of different programs \cite{gabriel2017effective}) \update{there are also inherent biases to which people can be prone to, unconsciously}{cognitive biases play a major role in humanitarian aid decisions~\cite{benartzi2001naive,bartels2006proportion,vastfjall2014compassion}. A charitable donor might struggle to decide how to split their personal donations among charities, or decision makers working on behalf of charities or governments may be confused about which programs should be given priority, when distributing the collected donations}. Many cognitive biases have been observed in prior research in cognitive and social psychology~\cite{daniel2017thinking}, especially when making a decision which requires the allocation of resources, like time or money \cite{null2011warm, bartels2011group, butts2019helping}. \update{For example, when presented with numbers which communicate the loss of life, people tend to feel more compassion towards a small number of individuals than a significantly large number \cite{butts2019helping}. This means that a donor willing to donate \$100 to 10 people in distress, might consider donating only \$110 when there are 100 who need aid, even if they have excess to donate.}{We identify that in the context of humanitarian resource allocation, prevalent cognitive biases stem from reasoning about people in need of aid as groups or communities, as opposed to considering each individual separately (\S \ref{section:cog_bias}). However, how to curb these biases remains an open question~\cite{kretz2018experimentally}.
}




\update{ 
We posit that visual analytics tools can be designed to aid people achieve more fair and equitable resource allocation decisions across multiple groups. Existing tools and visualizations primarily focus on encouraging their audience to contribute more towards a specific cause \cite{gabriel2017effective}. To this effect, some leverage knowledge in psychology, for instance, by using photography as a means to elicit empathy for people affected in tragedies~\cite{jenni1997explaining}. But few promote unbiased comparisons. Donation websites such as donorschoose.org or fundraising platforms like gofundme.com allow users to search for similar causes, but comparison across these causes is left to the user, without a guiding tool.
}{}
\update{}{
Motivated by the promise of visualizations to mitigate bias~\cite{dragicevic2022information, wall2019toward, dimara2018mitigating, koval2022you, ritchie2019lie}, we investigate how interactive visualization tools can be designed to aid people achieve more fair and equitable resource allocation decisions across multiple groups. More specifically, our first core research question is:}

\update{}{\rvis\ -- \textbf{Can visualization improve allocation fairness when compared to text?  }}

\noindent \update{}{Our work\update{also}{} expands on research that has studied the use of visualization for humanitarian purposes, which looked at eliciting donations~\cite{gabriel2017effective}, or prompting emotional connection to data \cite{morais2020showing}. Our study addresses other biases.
}

\update{}{We also explore the effect of a specific information framing strategy, i.e. showing resources allocated to individuals instead of showing resources allocated to groups.} Our research is grounded in past work looking at the effects of information framing on decision making tasks~\cite{kahneman1981simulation}, and nudging in computational interfaces~\cite{mota2020desiderata, cockburn2020framing}. These works suggest that the way information is presented can influence people's choices, but did not explore effects on fair allocations. \update{}{Our other core research questions are thus:}

\update{}{\rframe\ -- \textbf{Can presenting resource allocation information on a per-individual basis (showing the amount allocated to each person) improve fairness compared to doing so on a per-group basis (showing the amount allocated to each cause)?}}

\update{}{\rint\ -- \textbf{How do presentation format and information framing interact with one another?}}

We contribute the results of three crowdsourced experiments with large samples (N = 239, 244, 495) where participants were presented with different visual and textual framing in an interactive allocation tool used to allocate resources between two hypothetical charitable programs that are similar but provide support for populations of different sizes. Our main finding indicates that individual-framed visualizations and text may be able to curb unfair allocations caused by group-framed designs. Complementary findings are described in the sections about the experiments. This work opens new perspectives that can motivate research on how interactive tools and visualizations can be engineered to combat cognitive biases that lead to inequitable decisions.
\section{Related Work}\label{sec:related_work}
Our work draws from prior work studying cognitive biases, the effect of framing information in text and visualizations, and \update{tools}{tool designs relevant to} supporting resource allocation decisions.

\subsection{Cognitive Biases}
\label{section:cog_bias}

Efficient response to humanitarian crises requires pragmatic, unbiased decision making. But people can be prone to \update{a number of}{} biases which are well documented in psychology 
\update{}{(see comprehensive reviews~\cite{comes2016cognitive, caviola2021psychology})}. \update{This can have severe consequences. For instance, failure to allocate available resources fairly between charity programs and interventions aiding different communities may cause harm or even cost lives.}{A bias is a type of thinking error that is not necessarily caused by lack of knowledge or information, but rather by applying a decision rule that is helpful in some situations but not in others. In the context of charitable giving, cognitive biases can be harmful and may ``lead to the systematic misallocations of funds and waste of resources'' \cite{baron2011heuristics}.} 

\update{}{\textit{``If I look at the mass I will never act. If I look at the one, I will.''} This famous quote captures a
powerful phenomenon where the way people conceptualize individuals in need of help directly impacts their response to the issue. Our work is interested in this phenomenon and cognitive biases at play when allocating resources between charitable programs}. 

\update{}{The more the people in need (``the mass''), the less we feel compelled to aid}. The \emph{compassion fade} or \emph{compassion fatigue} effect is a cognitive bias that refers to a decrease in \update{helping or }{}compassionate intent and helping behaviour as the number of people needing aid increases \cite{butts2019helping, maier2015compassion, slovic2007affect, bhatia2021more}. For instance, \update{a donor}{one} might \update{give}{allocate} \$300 to a charity \update{}{program} supporting 10 children, but would not proportionally increase \update{their donation to}{this amount for} a \update{charity}{program} supporting 500 children\update{, which they may instead donate \$500 to}{}.
\update{}{The opposite is also true! The less the people we can help, the less we feel compelled to aid. The \emph{drop-in-the-bucket} effect refers to when people feel that helping few people accomplishes nothing \cite{bartels2011group}. Our work is interested in whether either of these effects generalizes to resource allocations, i.e. whether people tend to allocate resources to the larger or smaller group.}



\update{}{It could also be the case that neither apply. The \emph{diversification bias} effect states that people tend to allocate available resources evenly over different choices in contexts like stock investment or product consumption \cite{read1995diversification}, thereby discounting differences between the different options. Our work expects to identify whether this effect happens in the context of humanitarian resource allocation with visualizations.} 

\update{}{Recent research also suggests that the effect of \emph{entitativity}, which shows bias relative to perception of group belonging, can be another factor influencing reasoning~\cite{kogut2005identified}. For instance, children who are identified as part of a family are more likely to collectively receive more aid than children who are seen as a set of independent individuals without explicit group membership, or as statistics~\cite{smith2013more}. In the context of allocating resources between programs, we identify a tension between preserving entitativity of each group, and emphasizing individuality to focus on what amount of resources each person receives regardless of which program they are aided by.}

\update{}{Our study allows to evaluate whether tool designs are successful at curbing bias. We manipulate the number of people in two charitable programs, as well as the way information is presented (see \S\ref{section:framing}). This allows us to interpret results as to whether and why different resource allocation tool designs result in participants' resource allocation strategies that tend to favor the larger, smaller, or none of two groups, or whether individuals are equally served.}

\subsection{Framing Effects}\label{section:framing}
Tools that support decision making are concerned with the effective presentation of information relevant to the decision. We explore designs that\update{use textual information and visualizations. Each can}{} vary in the way information is framed, which research has shown can influence how people interpret information and make subsequent decisions -- a cognitive bias referred to as \textit{framing effect}~ \cite{kahneman1981simulation}. \update{}{Framing is a broad term, and there are numerous ways information can be framed. Framing effects can be referred to as any ``small changes in the presentation of an issue or an event, such as slight modifications of phrasing, [that can] produce measurable changes of opinion''~\cite{hullman2011visualization}. }



Text can be manipulated to communicate the same fact, but with a different framing. Take for instance a glass ``half full'' or ``half empty''. The underlying data is \update{technically}{}the same, but saying one or the other may influence how people interpret the same bit of information. Work by Kahneman and Tversky~\cite{kahneman1981simulation} showed that different framings about loss of human lives resulted in people making different decisions regarding actions they took. For instance, people are more sensitive to text saying: ``Of 600 people, 200 will be saved'' than they are to the statement ``Of 600 people, 400 people will die''. 



\update{}{Hullman and Diakopoulos~\cite{hullman2011visualization} propose a taxonomy of framings in the context of communicative visualizations, which covers changes to the data (e.g., aggregating data) and the way it is represented (e.g., using visual metaphors, using visual grouping).} 
\update{Researchers have explored framing applied to visualization design, and its influence}{Studies have explored the influence of visualization framing} on people's behaviors~\cite{hullman2011visualization}, interpretation of data~\cite{franconeri2021science, campbell2019feeling}, and trust in information~\cite{chevalier2013using}. A comprehensive review is beyond the scope of this paper, as it can be argued that the whole field is concerned with optimizing visualization design to achieve fair representation of data.

\update{}{Since we study how people split resources between two groups, we are interested in a specific framing effect: showing resources each \textit{group} gets as a result of the decision (group framing) vs. resources each \textit{individual} gets (individual framing). This focus stems from the nature of our research question and of the task itself: since cognitive biases covered in \S\ref{section:cog_bias} can be conceptualized as putting too much weight on entitative groups and not enough on individuals, we deem it important to look at whether individual framing vs. group framing can impact allocation fairness \rframe.}


\subsection{Design for Resource Allocations}
\label{section:vis}

Researchers have argued that computer tools can help overcome cognitive biases in decision making involving humanitarian scenarios~\cite{comes2015exploring, comes2016cognitive}. \update{}{However, we still lack comprehensive guidelines to avoid cognitive biases in decision making~\cite{kretz2018experimentally, wall2019toward}. To the best of our knowledge, there is no existing review of user interfaces that funding managers use in such scenarios, and research on developing interfaces to support unbiased resource allocations is also scarce}.

\update{}{Our study focuses on identifying if and how visualization can improve allocation fairness when compared to text \rvis{}. Allocated resources can be represented with simple visual media such as images or bar charts. However, studies have also focused on the design of the iconography and its influence on potentially aiding users to grasp concepts such as scale or help induce affective responses. Concrete scales~\cite{chevalier2013using} exploit visceral responses to the volume occupied by familiar objects re-expressing otherwise difficult-to-grasp quantities. Such representations can help better apprehend the amount of resources and people, and is a common strategy to raise awareness on casualties in a disaster (e.g. memorials using flags or shoes to represent victims). Anthropomorphized data graphics, which portray characteristics of humans, is another commonplace visual communication strategy employed as a potential means to elicit more prosocial behaviours \cite{boy2017showing, morais2021can}.
We note that conveying data through visual data storytelling to elicit empathy has had limited success, and prior studies do not indicate strong effects \cite{morais2021can}. Yet, these approaches may help create more relatable iconography when visualizing people in contexts where cognitive biases tend to put too much weight on entitative groups. Our design approach builds on this premise, but other designs could be explored in future work.}

\update{}{Prior studies have also investigated presentation and framing within interfaces and visualizations that help mitigate bias.
A tool allowing to remove irrelevant information locally within a visualization to make a comparative decision has been shown to mitigate biases such as the attraction effect \cite{dimara2018mitigating}. Visualizations which show predictions of future outcomes have been shown to encourage less optimistic estimates \cite{koval2022you}, and interaction with visualizations to present information from different perspectives has been proposed to help combat anchoring effects~\cite{ritchie2019lie}. Our work extends this line of research, by investigating the role of visualization and information framing on a set of cognitive biases unique to allocating resources in a humanitarian aid context (\S\ref{section:cog_bias}).}


\update{Other works have surfaced biases introduced by particular representations (e.g. error bars hindering inferences~\cite{correll2014error}). Researchers have also studied visualization techniques aiming at leveraging cognitive processes to a particular goal.

Past research also informs avenues for interactive techniques which can be applied in our context. Nudging, where incremental hints are made to favour certain outcomes, has been a method used in interface design to encourage more equitable donations across many different charities \cite{mota2020desiderata}.
}{}
Also relevant are tools that allow to explore the result of alternative scenarios, through direct manipulation and visualization.
\update{}{ Their design can serve as inspiration for features which allow for the user to visually compare the effects of different allocations.} For instance, the `What-if Tool' lets machine learning practitioners evaluate the effect of manipulating input data on machine learning models performance, such as ML fairness metrics~\cite{wexler2019if}. Through interactive widgets, users can change aspects of the datasets and model parameters, and see the results of such hypothetical situations. Similarly, tools like the climate change calculator~\cite{climatecalculator} allow viewers to visually explore the effect of varying government action scenarios on the future climate. Our tool borrows from these past works, by allowing users to dynamically adjust their allocation decision, and get immediate feedback on its result through visualization.

\section{A Tool for Resource Allocation: Problem and Design Considerations}
\label{section:framing_res_alloc}

Imagine that you decide to contribute to humanitarian causes -- how should you split your money if you had two different causes to aid? Or think about how organizations like the United Nations should allocate their funds between the programs they support. Those are examples of the \textbf{resource allocation problem}, which happens when two or more groups should be benefited by the allocation of resources. The question is: how to make effective or fair allocations based on the information that the decision-maker has available?

The problem is complex because, in real-world scenarios, groups can have completely different characteristics. \update{For the two tornado disasters from the introduction, we would have to compare an event that happened in Ohio with another that occurred in a small region in Nicaragua, which probably received less media coverage in Canada.}{ } Those factors can play a role in how people make judgements to allocate money. If people rely on their own experience or intuition, they might be influenced by cognitive biases and their attitudes. On the other hand, if decision makers decide to allocate resources based on data but without a proper tool, they can be overwhelmed by the amount of information.

This paper is a first attempt to explore the design of resource allocation tools to promote fair allocations. We consider resources to be \textit{fairly allocated} when each beneficiary receives the same amount of resources as any other beneficiary, irrespective of the group they belong to. For the sake of simplicity and reproducibility, we focus on a task in which all individuals from two groups have comparable needs and contexts, with only the group sizes changing. This way, we make sure that the same amount of resources would benefit different individuals equally. This decision helps breaking the resource allocation problem down into a simpler problem and controls for confounding factors in the experiments we conducted. Possible limitations for this approach are addressed in \S\ref{subsec:limitations}.

We are specifically interested in the effect of two design dimensions on resource allocation. \textbf{Presentation format} corresponds to how the resource allocation information is represented  -- for example, we can use \textit{text} to describe the resource split across groups, or a \textit{visualization}. \textbf{Information framing} refers to how precisely the allocation information is conveyed to the user -- for example, a \textit{group} framing describes the amount donated to each group, whereas an \textit{individual} framing focuses on how much would be allocated to each individual. We examined how presentation format \rvis and information framing \rframe affect resource allocation by designing and testing interactive tools that vary in terms of those dimensions. In the following, we describe examples of designs that combine different presentation formats and information framings, and introduce the interactive tools used in our experiments.

\subsection{Textual Information}\label{sec:text}

Information needed to make resource allocation can often be composed of a textual description of a problem accompanied by other presentation formats such as a picture or a visualization. The way the text is framed can affect people's judgements~\cite{chang2009framing}. We explore two ways of framing the textual information as follows:

When \textbf{textual information} is \textbf{framed by group}, the message is expressed as a total donation to each group (e.g., \textit{"give \$1,000 to aid the victims of Florida's hurricane and \$300 to help victims of Haiti's disaster"}). This framing allows readers to think holistically about the allocation given to the respective groups. 
This can induce the effect of entitativity \cite{smith2013more} as it can encourage the donor to associate the group of people in each program to singular group units, possibly resulting in compassion fade~\cite{slovic2007affect} or drop-in-the-bucket~\cite{bartels2011group}. Group framing can also result in the naive diversification bias~\cite{benartzi2001naive}, when people prefer to allocate resources evenly across groups, even when these have different characteristics.

When \textbf{textual information} is \textbf{framed by individuals}, the message is expressed in terms of the amount received per person (e.g., \textit{"give \$15 to each victim of Florida's hurricane and \$12 to each victim of Haiti's disaster"}). This framing allows readers to reason in terms of what individuals in the respective communities each receive. 
This might alleviate the entivativity and diversification biases, since individuality is directly expressed, as well as the effect of compassion fade, because large numbers become more palatable when expressed as what that means for individuals.

\subsection{Visual Information}
\label{sec:visualization}

\begin{figure*}[t]
        \centering
        \includegraphics[width=0.7\textwidth]{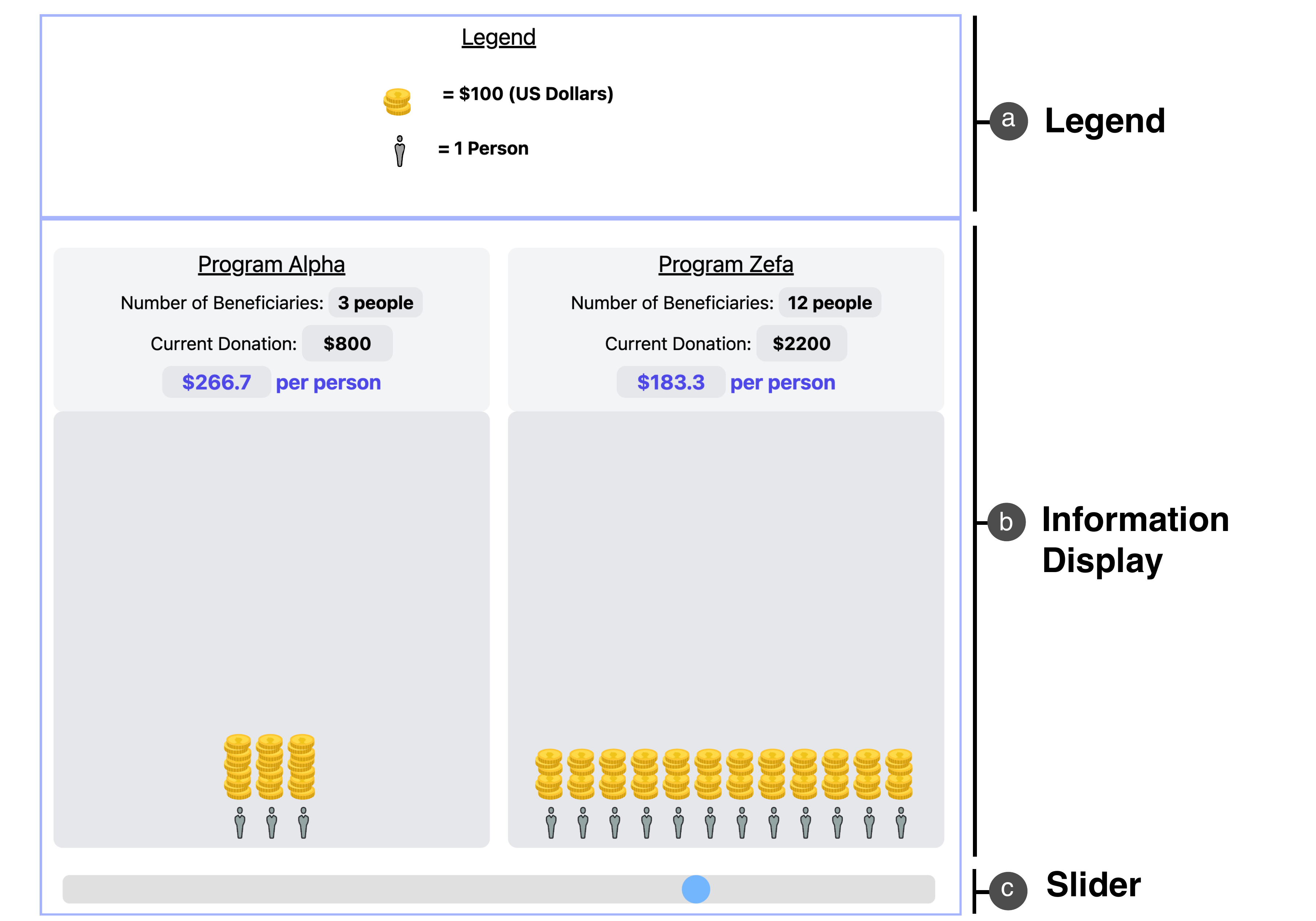}
        \caption[]
        {Screenshot of our instance of an interactive resource allocation tool featuring two charitable programs, used in Experiment 1 of our study. With the help of the legend (a), a donor can directly assess the result of resource allocation presented in the information display (b). Here both text and visualization representations are displayed, each presenting allocations with an individual framing. The donor can dynamically adjust how the resources are split between the two programs, using the slider (c), which automatically updates the information presented in the information display.  }    
        \label{fig:interface}
        \Description{Three photos labeled (a), (b) and (c) show pictures of an interactive visualization. Image (a) shows a legend. The legend a stack of three coins equal to 100 US dollars and a human silhouette equal to 1 person. In (b) it shows an information display with two boxes beside each other representing information about a program. At the top of each box is the name of text description with of the number of beneficiaries, current donation amount and money given per beneficiary. At the bottom of each box are human silhouettes with stacked coins on top. In (c) there is a slider with a blue handle for the slider-thumb.}
\end{figure*}

For the display of information using a data visualization, we opt for an iconic representation~\cite{boy2017showing} of victims (i.e., icons of a human), and resources (i.e., coins). Like for Isotypes~\cite{neurath1974isotype}, the number of icons directly encodes quantities, using partial rendering of icons when necessary (see ~\autoref{fig:interface}). \update{We chose an iconic representation to represent individuals as the true scale of the communities may be easier to grasp through visual aggregation of icons, than by interpreting numbers in textual form.}{} To vary framing, we manipulate the organization of icons representing quantities as follows:

\textbf{Visual information} can be \textbf{framed by group} using Gestalt psychology~\cite{kohler1967gestalt} to create a perception of unity for each group. Icons representing people from the same groups can be located together so that they are perceptually perceived as a unit. Similarly, icons representing resources can be grouped together spatially to be interpreted as an overall amount of resources allocated to the group. We chose spatial grouping (see an example in \autoref{fig:exp_1_conditions}-c), but other perceptual groupings could be considered in future designs.

\textbf{Visual information} can be \textbf{framed by individuals} using the same Gestalt principles, combining visual representations that correspond to each beneficiary. An individual framing can be created by placing the exact amount of resources allocated to each individual, besides each individual (see an example in \autoref{fig:exp_1_conditions}-d). This way, viewers can easily perceive the result of splitting one group's allocation across individuals in that group.

\update{}{We used an iconic representation to show individual people because it makes it easier for viewers to understand the true scale of a community.}
\update{}{Although it is possible to use aggregated visual encodings such as bar charts to convey population sizes, such an aggregated representation would not work well in an individual framing, where the goal is to emphasize the share of resources each individual gets. In contrast, a unit chart representation where each person is represented by a symbol works well with both group and individual framings. Unit charts are also commonplace when conveying information about populations, such as in data journalism~\cite{morais2020showing}.}

\subsection{Putting it all together: An interactive tool for resource allocation}
\label{sec:tool}

We created a resource allocation tool that combines different presentation formats and framings. \autoref{fig:interface} shows a screenshot of the interface we used in our study. The tool encompasses: (\texttt{a}) a legend, which informs the number of people and resources encoded by each mark in the visualization; (\texttt{b}) a display, showing textual and/or visual information about the groups and corresponding allocations; and (\texttt{c}) a slider, which allows the user to change the amount of resources to each group. Since we only consider 2-group allocations, the display is divided in two parts; one for each program/group. The display dynamically updates when the user moves the slider to the left (to benefit group Alpha) or to the right (to benefit group Zefa). That template was used for all of our experiments.

To study how the two dimensions affect resource allocation, we manipulated the tool design in terms of presentation format and information framing. We tested designs composed of only text, visualization, 
or both. Also text and visualization had group or individual framings. In the example shown in \autoref{fig:interface}, the display encompasses both textual and visual information. The text follows an individual framing because it contains the exact amount to be donated for each individual (in blue). The visualization also follows an individual framing because the coins encoding the values allocated to each beneficiary are aligned with that representing individuals. As both text and visualization have the same framing, we consider that the tool has a \textbf{congruent framing}. Conversely, when the text and visualization have opposite framings, the tool has an \textbf{incongruent framing}. There are many other aspects to the design of a resource allocation tool that could be considered, which we cover in the discussion section. 

\section{Experimental Design: Overview}\label{sec:expe-design}

We conducted three experiments to gain a better understanding of how people approach the resource allocation problem in a fictional scenario. Each experiment builds on findings and lessons from the previous experiment(s). We describe the general context for our study and overall procedures. Elements specific to a particular experiment are discussed in the corresponding sections. All experimental materials and scripts 
were pre-registrered for each experiment.\footnote{All pre-registrations can be found at \url{https://osf.io/xqser?view_only=1049a79a474a4310958571269aa5e248} (Experiment 1), \url{https://osf.io/7ebsm/?view_only=4ff3f6ae865f444282ecd95de4a302b0} (Experiment 2) and \url{https://osf.io/egv85/?view_only=e702e7de07114e18bebd175aa0d8b43b} (Experiment 3). General supplementary material can be found at \url{https://osf.io/zrbej/?view_only=90e4c3581ece485690a39c487c0fb67c}.}

\subsection{Task}
\label{sec:task}
We created a fictional scenario for the resource allocation task. Participants were provided with a brief about a charitable organization which is helping setting two separate programs for impoverished children. The scenario brief specifies that the charity's mission is to aid children, and that the money to allocate will be of significant value to accomplish this goal, followed by instructions on the task:
\begin{quote}
\textit{``Your task is help the organization decide how to distribute money between two of their programs.''} 
\end{quote}

\noindent The two programs, Alpha and Zefa, differ in size:  Alpha aids less children than Zefa does (sizes are reported for each experiment). The brief specifies how the money will be utilized -- \textit{`` the money will go toward food, clothes, and education''}, but it includes little information about individuals, as we wanted to measure the presentation of summarized information rather than the presentation of identifiable victims (which comes with biases outside our scope~\cite{jenni1997explaining, friedrich2010individual}).

Our scenario was inspired by the scenarios found in related psychological studies
~\cite{garinther2022victim}. \update{}{We chose money because it is the main resource that decision makers deal with in charitable giving contexts, and monetary donations is common in similar studies from psychology~\cite{vastfjall2014compassion}, economics~\cite{schons2017should}, and HCI~\cite{boy2017showing,morais2021can}.} We chose to keep population sizes relatively small, i.e not in the thousands or millions, since a smaller number can be easier for participants to reason about both when shown in text and when encoded in a visualization. We believe this is a reasonable choice as similar studies showed that compassion fade stems from the proportional difference between populations \cite{maier2015compassion}, meaning that effects observed on small populations would theoretically hold for population sizes tens or hundreds times larger \cite{garinther2022victim}.

\subsection{Procedure}
\label{sec:procedure}
We deployed our experiment online. Upon sign up, participants were directed to a self-hosted website where they were first prompted for consent. Participants who consented were presented with the scenario and task (see \S \ref{sec:task}). When proceeding to the next screen, they were presented with the interactive tool to make their allocation (see \S \ref{sec:tool}). Participants were not instructed anything more than allocating resources between the two programs as they saw fit. No indication of performance goals was provided, i.e. participants were not hinted at what would be a reasonable or desirable split of resources, neither were they given time or speed constraints.

After submitting their chosen allocation, participants were asked to explain their rationale in a text box. We did not inform participants that they were going to be asked to justify their decision until after they decided how to split the resources between the programs, and participants were not allowed to return to the tool to change their allocation. We did so to minimize potential social desirability bias~\cite{fisher1993social} which may arise when one feels they will be held accountable for their decision, which is also why we opted against a donation intention question for our task. Participants were also asked to rate confidence in their choice, on a five-point Likert scale (``not at all confident'' to ``very confident''). 

Finally, participants were posed a multiple-choice question about the purpose of the charity, which we used as an attention check to exclude potentially inattentive participants or bots, whose responses would introduce noise.

\subsection{Measurements}
\label{sec:measurements}
Our primary measurement is obtained from the slider (\autoref{fig:interface}-c), which records how the full amount of available resource was split between the two programs. We 
developed metrics to gauge the extent to which allocations were \textbf{fair} (or unfair), described in each Experiment's section. We also characterized what different allocation values mean in terms of \textbf{allocation strategy} employed, as follows:

\begin{itemize}
    \item The \emph{\textbf{fair allocation}} strategy treats all children equally, i.e. money is allocated such that individual people in the two programs get the same amount of money. This allocation favors the larger program (Zefa), because more money is allocated to it compared to Alpha, but this strategy treats individuals equally, which can be conceptualized as the allocation strategy that is fair towards individuals. 
    \item The \emph{\textbf{unfair allocation}} strategy encompasses any other allocation, where children in different programs are aided a different amount of resources.
\end{itemize}

\noindent We further break down the \emph{\textbf{unfair allocation}} strategy into four subcategories as follows:

\begin{itemize}
\item The \emph{\textbf{minority-skewed allocation}} strategy favors the program with less people, i.e., more money is allocated to the program with the smaller number of children. The allocation favors both the smaller program (Alpha) and the individual people in this program.

\item The \emph{\textbf{naive allocation}} strategy treats the programs equally, i.e., the same amount of money is allocated to the two charity programs. The allocation treats charity programs equally but favors individual people in the smaller program (Alpha). This response is indicative of diversification bias \cite{benartzi2001naive}. 

\item The \emph{\textbf{mixed allocation}} strategy balances causes and beneficiaries, i.e., more money is allocated to the program with the larger number of children (Zefa), but individual people from the smaller program (Alpha) still get more money. 
This type of response is indicative of compassion fade effect \cite{butts2019helping}.

\item The \emph{\textbf{majority-skewed allocation}} strategy favors the program with more children, i.e., more money is allocated to individual people in the charity program with the larger number of children (Zefa). The allocation favors both the larger program and the individual people in this program, and indicates a drop-in-the-bucket bias~\cite{bartels2011group}.

\end{itemize}



\subsection{Analyses}
\label{sec:analyses}

\update{}{We follow an estimation approach to statistical reporting, which focuses on reporting effect sizes and interval estimates instead of p-values, and encourages non-directional research questions \cite{cumming2013understanding,dragicevic2016fair}.} We draw our primary inferences from graphically-reported point and interval estimates \cite{cumming2005inference}. Effects are estimated using the percentile bootstrap method to obtain 95\% confidence intervals (CIs). We distinguish between planned (preregistered) and post-hoc analyses. Instead of adjusting for multiplicity, we distinguish between primary and secondary research questions \cite{bender2001adjusting, li2017introduction}. We also performed a qualitative analysis of the content of the free-text responses to the question asking participants to justify their choice. Methods used are described for each Experiment.

\subsection{Recruitment, Inclusion \& Exclusion Criteria}
Participants were recruited through the Prolific crowdsourcing platform, which was chosen for its larger set of positive actors compared to other similar platforms \cite{gupta2021experimenters}. In a pre-screening process, participants were required to speak English fluently, to have a 99\% approval rate on Prolific, and to have not participated in any of our pilots to be eligible. We also recruited only from majority English speaking countries (UK, USA, Canada). We excluded data from participants who did not complete the study, or failed at the attention test (see \S \ref{sec:procedure}).

\subsection{Implementation}

\update{}{
     We deployed our study tool as a standalone web-based page compatible with most browsers. Our tool is built in Typescript 
     with React.js, D3.js, and Tailwind.css libraries for the frontend, and Node.js for the backend.
     Besides practical considerations for a smooth crowdsourced study experience, we also chose a web-based implementation as future extensions of our tool may be integrated into larger websites such as donorschoose.org or gofundme.com. 
}

\section{Experiment 1: Initial Investigation of Average Resource Allocation}

Since no prior studies, to our knowledge, have looked at visualization framing in the context of charitable resource allocations, we designed this first experiment to investigate which donation strategies people follow when presented with text and visualizations with different framings. We also set out to verify whether effects caused by cognitive biases \update{such as drop in the bucket \cite{friedrich2010individual}, compassion fade \cite{butts2019helping}, and \update{}{diversification ~\cite{read1995diversification}}{(\autoref{section:cog_bias}}} would manifest within our study design. More importantly, we explored the effect of presentation format and information framing on resource allocation. 

\update{}{In our pre-registrations, we distinguish between our primary (or main) research questions, and secondary (or auxiliary) research questions. Each primary research question further elaborates on one of our core research questions outlined in the introduction, and is indicated in parenthesis.}

\noindent Our research questions were:
\begin{enumerate}[itemsep=0pt, topsep=0pt, leftmargin=6em]
    \item[\update{}{\rvis\ } \textbf{\small RQ1.1}] To what extent does a visual representation of information affect donation allocations compared to a design with the same information presented as text only?
    \item[\update{}{\rframe\ } \textbf{\small RQ1.2}] To what extent does framing the donation amount per individual affect donation allocations when compared to framing the donation amount per program?
    \item[\update{}{\rint\ } \textbf{\small RQ1.3}] To what extent do the factors of information representation and information framing interact with one another regarding donation allocations?
    \item[\textbf{\small RQ1.4}] What are the most common stated reasons for the donation allocations?
\end{enumerate}

\subsection{Experimental Design}
Experiment 1 followed the procedure described in \autoref{sec:expe-design}. The Alpha and Zefa programs aided 3 and 12 children respectively. The amount to allocate was USD\$3,000.


\begin{figure*}[t]
    \centering
    \includegraphics[width=\textwidth]{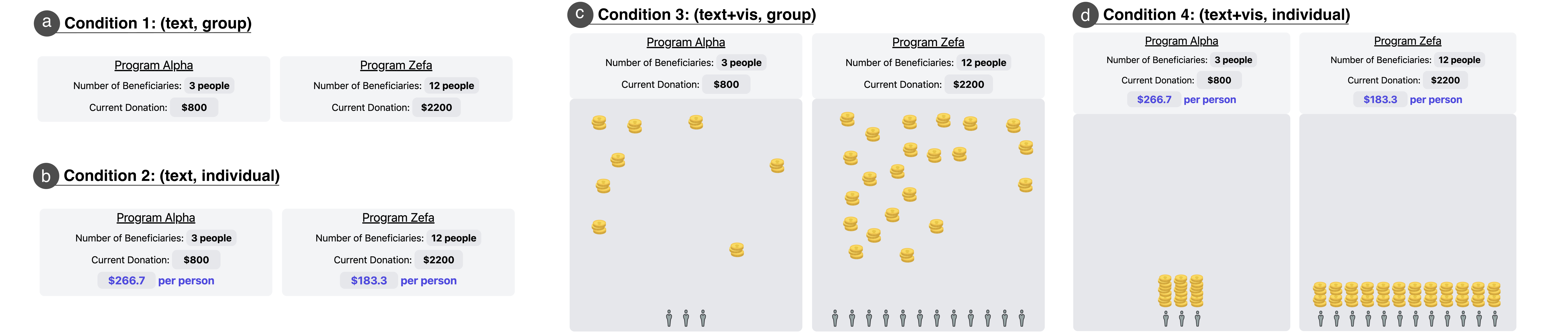}
    \caption[]{Experiment 1: Stimulus used in the different conditions. See a demo of the tool at \url{https://resallocdesign.github.io/}}
    \label{fig:exp_1_conditions}
    \Description{Three photos labeled (a), (b), (c) and (d) show pictures of visualizations or text. Image (a) shows shows an information display with two boxes beside each other representing information about a program. At the top of each box is the name of the program, the number of beneficiaries and current donation. For example, "Program Alpha", "Number of Beneficiaries: 3 people", "Current Donation: \$600". In (b) it shows the image in (a) but with money given per beneficiary as well, i.e "\$266.7 per person". In (c), there are two boxes. The text at the top of the box is the same as (a) but at the bottom of each box are human silhouettes with coins scattered around the box. In (d), there are two boxes as well. The text at the top of each box is the same as in (b) but at the bottom of each box are human silhouettes with coins stacked on top of each individual silhouette. }
\end{figure*}

Our between-subject design manipulated \update{}{two independent variables. The} \textsc{presentation-format} \update{--}{ determined whether} the same information was either conveyed using using text only (\textit{text}), or  \update{}{using} both text and visualization (\textit{text+vis})\update{; and}{.} \update{}{The} \textsc{framing} \update{of both of }{ variable determined whether}  \update{these}{the} presentation formats (\S\ref{sec:tool}) \update{-- the same information was presented}{present information} with either a focus on splitting resources between causes (i.e. \textit{group} framing), or a focus on splitting resources between individuals in the causes (i.e. \textit{individual} framing). Our full factorial design encompassed four conditions. The content presented in the display component of the tool (see \autoref{fig:interface}-b) for each condition is depicted in \autoref{fig:exp_1_conditions}.

\subsection{Participants}
Our target sample size was 256 participants ($\sim$64 per condition; participants were randomly assigned a condition at sign up). This gave us a 0.8 power to detect a ``medium'' Cohen d’s effect size of 0.5 between any two conditions (per the G*Power software) for differences between independent means. \update{}{We ran this study on August 7, 2021 until target sample size was reached (within hours), after 2 pilots with 5 and 10 participants respectively}. Out of the 256 participants who consented, 17 were excluded for not completing the study or failing the attention check. The 239 remaining responses were divided as follows: 63 were assigned the \emph{(text+vis, individual)} condition, 63 completed the \emph{(text, individual)} condition, 56 completed the \emph{(text+vis, group)} condition, and 60 completed the \emph{(text, group)} condition.

\subsection{Measurements \& Analyses}
\label{sec:exp1-analysis}
We followed methods described in \S\ref{sec:measurements} \& \S\ref{sec:analyses}. We used the allocation to the larger group (Zefa) to measure allocation strategy: a value of \$2,400 corresponds to a fair allocation strategy, where all individuals are given the same amount. \update{}{With this measure, we used a difference in means to generate the bootstrapped confidence intervals used to answer our research questions.}

For the qualitative coding of justification comments (method not preregistered), we followed an emergent coding approach \cite{stemler2000overview}: four coders independently assigned labels to participant responses. One coder did so across all responses while the three other coders each covered a separate one-third of the responses. Amount allocated to each program was hidden during this process. All coders discussed respective codes, refining, combining, and merging codes where appropriate, converging on a common consolidated coding scheme comprising of 5 categories, each of which comprising several codes (see Supplemental). One coder then re-coded all shuffled responses using the new scheme.

\update{}{
  Our approach to data collection and methods for analysis were refined through unregistered piloting of our study. This helped us refined our wording and layout of the story, adding an attentional check, and refining analysis scripts.
}

\begin{figure*}[t]
    \centering
    \includegraphics[width=\textwidth]{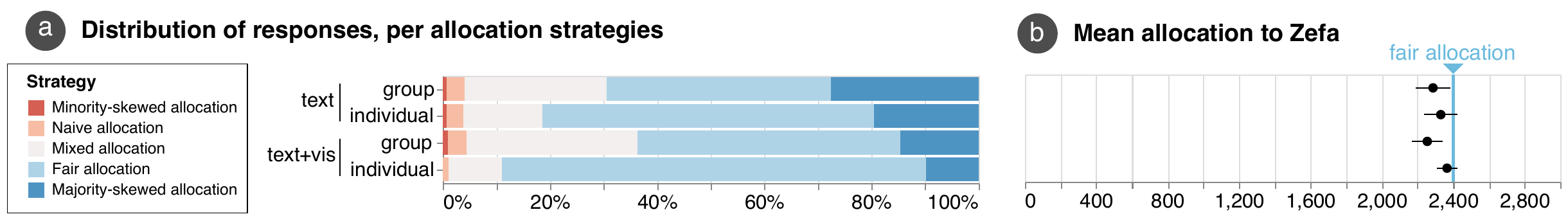}
    \caption[]
    {Results of Experiment 1: (a) Distribution of allocation strategy chosen by participants, by \textsc{Presentation Format} and \textsc{Framing}. (b) Mean allocation granted to Zefa by condition. Error bars are 95\% CIs.}
    \label{fig:exp_1_summary}
    \Description{There are two plots (a) and (b) which describe the distribution of responses per allocation strategy. Plot (a) shows a stacked bar chart broken down by five categories. The 5 categories given in the legened are: "minority-skewed allocation", "naive allocation", "mixed allocation", "fair allocation", "majority-skewed allocation". There are four condition described in Experiment 1. The fair allocation strategy is generally has the highest proportion across all conditions, around 40-80\%. The group condition has around 10\% lower fair allocation that individual framed conditions. Plot (b) shows the bootstrapped confidence intervals to the mean allocation given to Zefa, across all conditions. All confidence interval range from around 2200 to 2400. Confidence intervals for individual-framed conditions cross 2400. All mean point estimates are roughly in the center of all intervals.}
\end{figure*}

\subsection{Results}
\autoref{fig:exp_1_summary} shows a summary of results. The stacked bar chart (\texttt{a}) shows a break down of responses per allocation strategy (see \S \ref{sec:analyses}), and the error bars (\texttt{b}) show 95\% CIs of mean allocation to Zefa, compared to a fair allocation (\$2,400).

Overall, most participants followed a fair allocation strategy. \autoref{fig:exp_1_summary}-b shows mean allocations to Zefa nearing or covering the \$2,400 value. \autoref{fig:exp_1_summary}-a shows that the condition with fewer participants choosing a fair allocation (in light blue) was the one with textual information and group framing (40\% of participants), while showing text+visualization with individual framing led to about 80\% fair allocations. Also, a notable amount of participants chose a mixed allocation strategy in \textit{group} framing conditions (30\% for \textit{text}, 34\% for \textit{text+vis}), which shifted the mean allocations of those conditions farther from the fair allocation. Below, we discuss our results organized by research question.

\update{}{\rvis} \textbf{RQ1.1: Effect of Adding a Visualization Support.} We were interested in learning the extent to which different presentation formats affect resource allocation. \autoref{fig:exp1-detailed-results}-a shows the effect of supplementing text with a visualization on our metric, by framing. We could not find evidence that the addition of a visualization affects the fairness of the mean allocation compared to a design with the same information presented as text only.

\update{}{\rframe} \textbf{RQ1.2: Effect of Framing.} We also explored the extent to which different framings affect resource allocation. \autoref{fig:exp1-detailed-results}-b shows the effect of varying framing on our metric, for each presentation format. Overall, we have some evidence suggesting that an individual framing leads to a fairer mean resource allocation than a group framing, for the condition where both textual and visual information are available.

\begin{figure*}[t]
    \centering
    \includegraphics[width=\textwidth]{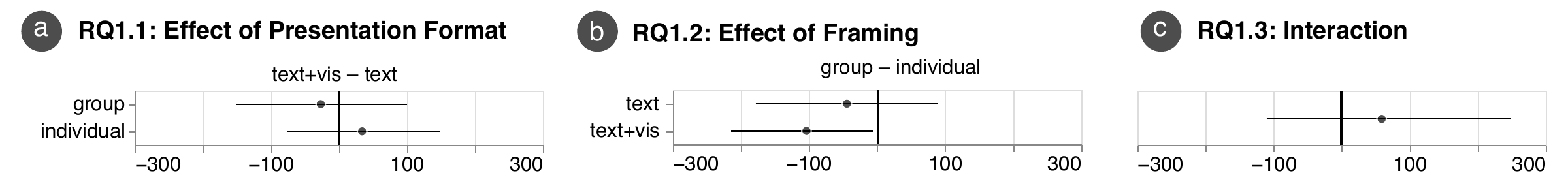}
    \vspace*{-0.7em}
    \caption[]
    {Results of Experiment 1: Effect of \textsc{Presentation Format} (a) and \textsc{Framing} (b) on the mean allocation (here in \$), and interaction between our different factors (c). Error bars are 95\% CIs.}
    \vspace*{-0.7em}
    \label{fig:exp1-detailed-results}
     \Description{There are three plots (a), (b), and (c) which show the effect sizes using the difference of means confidence intervals for the distribution of responses per allocation strategy. Plot (a) shows the difference of means between the text+vis minus text variables, across the group and individual framing conditions. Both intervals are between -150 to 150 and both beans are roughly close to zero. The CI for the group condition is the left of zero while the individual condition is to the right of zero. Plot (b) shows the difference of means between the group minus individual framing variables, across the text and text+vis condition. The both intervals are between -170 to 100. The condition where text+where is present, the interval does not cross 0. Plot (c) shows the confidence interval for the interaction value, and is ranged from around -100 to -250. }
\end{figure*}

\update{}{\rint} \textbf{RQ1.3: Interaction.} We investigated the interaction between the two design factors (i.e. presentation format and information framing). \autoref{fig:exp1-detailed-results}-c shows a quantification of this interaction. We could not find evidence that our factors interact with respect to mean resource allocation. 

\textbf{RQ1.4: Stated Rationale.} We also looked for the most common stated reasons participants provide for their donation allocations. From coding the justifications (see \S \ref{sec:exp1-analysis}), our main insights are:

(i) Inconsistencies between the strategy used and the justification provided. The \emph{sense of equality} code determined if participants responses indicated that they tried to allocate equitably across charitable programs. We coded 70 responses as such (out of 239), however 12 of those participants had not used a fair strategy.  

(ii) Mental calculations (e.g. dividing a total amount by number of individuals) are not necessarily performed to achieve a fair allocation, and are prone to errors. The \emph{math} code category indicated whether the response included some explicit calculation. Of the 44 participants who did include such mention, eight had not used the fair allocation strategy.

(iii) Some participants made assumptions about the programs, which have likely impacted their response. The \emph{contextual influence} code category determined whether the response indicated that the participant's allocation was chosen based on personal interpretations of the scenario, e.g. presuming that program Zefa solves a more significant problem, whereas our scenario did not state so. This finding motivated us to reword the scenarios in subsequent experiments, to leave less room to such extrapolations. 

(iv) The two other coding categories did not provide additional insight, and are thus omitted in interest of space.

\subsection{Discussion}

Contrary to our expectations, presentation format does not seem to greatly affect resource allocations \rvis. We could not find evidence that supplementing text with a visualization produces fairer mean resource allocation compared to only presenting text. This is inconsistent with our initial assumption that showing a visual representation would help participants to evenly distribute resources across beneficiaries. Note that we did not find that visualization negatively affects allocation fairness either.

Our results also suggest that the way the information is framed could affect how people allocate resources \rframe. We specifically found evidence that individual framing influenced participants to perform fairer allocations compared to a group framing, in presence of a visualization. We suspect that presenting the exact amount that each individual would receive influenced participants to choose a value that would help everybody equally. This finding reinforces previous evidence that information framing can play a role on decision making \update{\cite{cockburn2020framing}}{\cite{cockburn2020framing, gosling2020interplay}}.


We later realized that the weak evidence for the effect of framing could be attributed to the aggregated metric that we used. Mean allocation and mean differences are sensitive to extreme values; and the majority-skewed responses could have pushed the mean closer to 2400, even in conditions where a smaller proportion of participants made fair allocations. We decided to re-analyze our data using the proportion of fair allocations -- the ratio of fair allocations to the number of responses in each condition. This new metric better aligns with our research aim of exploring fair allocations. We did find evidence that individual framing causes more participants to use a fair allocation strategy (which can also be seen in the descriptive statistics of \autoref{fig:exp_2_summary}-a). Since this unplanned analysis was not preregistered, we re-ran a new experiment (Experiment 2) to confirm our tentative findings with this alternative metric.

\section{Experiment 2: A Closer Look at Fair vs. Unfair Allocations}

This second experiment is a replication of Experiment 1 with a more robust and meaningful metric, and improvements to the scenario brief based on limitations observed in Experiment 1. Our research questions were:

\begin{enumerate}[itemsep=0pt, topsep=0pt, leftmargin=6em]
    \item[\update{}{\rvis\ } \textbf{\small RQ2.1}] To what extent does a visual representation of information influence the proportion of fair allocations compared to a design with the same information presented as text only?
    \item[\update{}{\rframe} \textbf{\small RQ2.2}] To what extent does framing the donation amount per individual influence the proportion of fair allocations when compared to framing the donation amount per program?
\end{enumerate}

\subsection{Experimental Design}
We followed the same procedure and design as Experiment 1. The only difference was in the scenario brief. To prevent participants from speculating on the programs' effectiveness, we added: (i) a description of needs (i.e., \textit{``Children from both programs are in a very similar situation and need money equally. They will not benefit from any other donation than yours.''}), and (ii) a clarification of how the money will be allocated within the programs (i.e. \textit{``All the money you donate to a program will be split evenly between the children in the program.''}). See a demo at {\small \url{https://resallocdesign.github.io/}}. With these clarifications about the underlying assumptions, there is less ambiguity about which allocation is the most fair.

\subsection{Participants}
Like for Experiment 1, we targeted a sample size of 256 ($\sim$ 64 per condition). \update{}{We ran this study on March 1, 2022 until target sample size was reached.} Twelve (12) participants were excluded. The 244 remaining responses were divided as follows: 62 in the \emph{(text+vis, individual)} condition, 63 in the \emph{(text, individual)} condition, 54 in the \emph{(text+vis, group)} condition, and 65 in the \emph{(text, group)} condition.

\begin{figure*}[t]
    \centering
    \includegraphics[width=\textwidth]{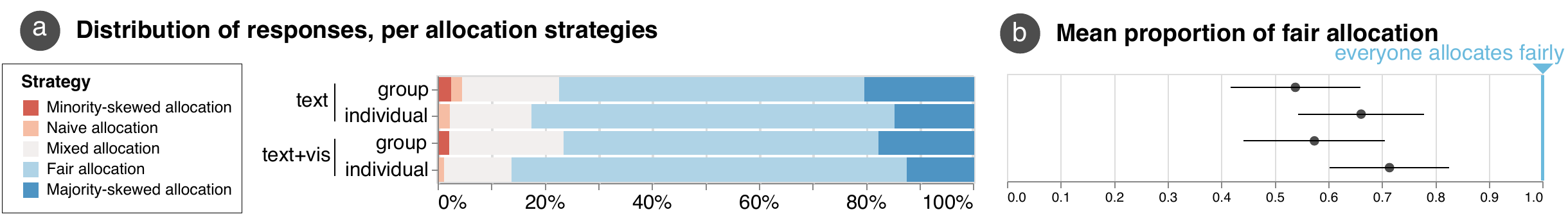}
    \caption[]
    {Results of Experiment 2: (a) Distribution of allocation strategy chosen by participants, by \textsc{Presentation Format} and \textsc{Framing} (see \S \ref{sec:tool}). (b) Mean proportion of people making a fair allocation by condition. Error bars are 95\% CIs.}
    \label{fig:exp_2_summary}
    \Description{There are two plots (a) and (b) which describe the distribution of proportion of people making a fair allocation by condition. Plot (a) shows a stacked bar chart broken down by five categories described in Experiment 1. The fair allocation strategy is generally has the highest proportion across all conditions, around 55-75\%. Plot (b) shows the bootstrapped confidence intervals and point estimates of proportion of people making a fair allocation by a specific condition. The first and third one range from around 0.4 to 0.7 with second and fourth range from around 0.55 to 0.83. Point estimated are at the center of each interval.}
\end{figure*}

\subsection{Measurements \& Analyses}
Unlike in Experiment 1, here we used the relative proportion of people who chose to use a fair allocation strategy as a means to assess overall allocation fairness.
While we report descriptive statistics showing the breakdown of responses per allocation strategies using all five strategies from \S\ref{sec:measurements}, our inferential statistics only contrast fair vs. unfair responses. \update{}{With this measure, we used a difference in means to generate the bootstrapped confidence intervals.}

\update{}{We validated our analysis methodology through unregistered experimentation: we explored the data from Experiment 1 to arrive at our measure used for analysis in Experiment 2. All analyses in Experiment 2 were pre-registered.
}

\subsection{Results}
\autoref{fig:exp_2_summary} shows a summary of our results. The stacked chart \texttt{(a)} shows responses per allocation strategy, by condition. The plot \texttt{(b)} shows the proportion of fair allocations.

Both plots suggest that most participants performed a fair allocation. We observe that individual-framing conditions tend to result in slightly higher proportions of fair allocations compared to group-framing ones. Actual evidence for these possible effects is further discussed under our research questions below.

\begin{figure*}[t]
    \centering
    \includegraphics[width=0.8\textwidth]{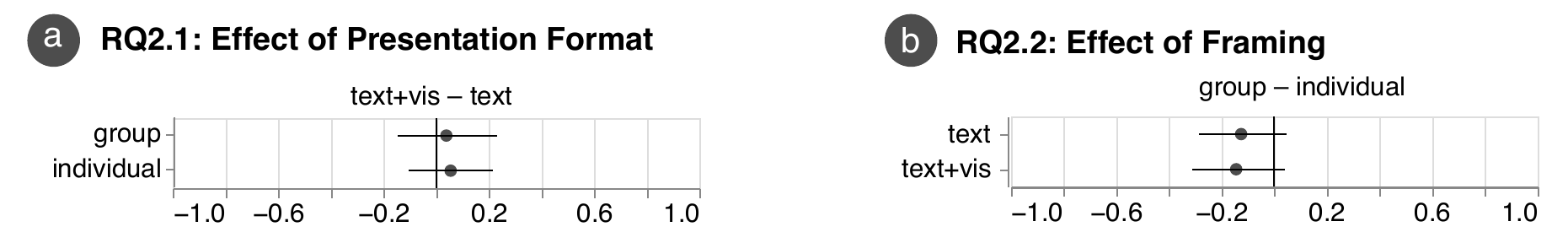}
    \caption[]
    {Results of Experiment 2: Effect of \textsc{Presentation Format} (a) and \textsc{Framing} (b) on overall proportion of fair allocations made. Error bars are 95\% CIs.}
    \label{fig:exp_2_analysis}
    \Description{Two plots of confidence intervals and point estimates labelled (a) and (b). Plot (a) shows two intervals slightly to the right of zero. Both are similar to one another and in between -0.2 and 0.2. Mean is close to 0. Plot (b) shows two intervals to the left of zero. Both are around -0.4 and 0.1 and the mean for both is around -0.12.}
\end{figure*}

\update{}{\rvis\ } \textbf{RQ2.1: Effect of Adding a Visualization Support.}
We were interested in the effect of presentation format on the proportion of fair allocations. \autoref{fig:exp_2_analysis}-a shows the difference in proportions between the \textit{text} format and the \textit{text+vis} format for each of the two framings. The effect of format seems to be negligible, independently of framing. 

\update{}{\rframe\ } \textbf{ RQ2.2: Effect of Framing.}
We were also interested in the effect of information framing. \autoref{fig:exp_2_analysis}-b shows the difference in proportion of fair allocations between \textit{individual} and \textit{group} framing, for each presentation format. There is evidence that individual framing may motivate fairer allocations than group framing: taken individually each of the two CIs provides very weak evidence, but taken together they provide converging evidence.

\subsection{Discussion}

The findings from this experiment are in line with the results from Experiment 1. 
\update{First, }{This means for our core question \rvis,} results suggest that supplementing text with a visualization produces a comparable proportion of fair allocations as using stand-alone text. 
\update{Second,}{For our core question \rframe,} we found some evidence that individual-focused framing is more effective for motivating fair allocations than group framing (with weak evidence for each of the two presentation formats). Those results strengthen the assumption that information framing plays a role on resource allocation. 

The high variation of fair allocations across conditions in the previous experiment was not reproduced in this experiment (compare \autoref{fig:exp_2_summary}-a and \autoref{fig:exp_1_summary}-a). As previously mentioned, maybe some instructions in Experiment 1 led participants to believe that a specific program was more important than the other, which caused a greater variance in responses. The instruction improvements in this experiment may have reduced room for personal assumptions and made the fair allocation strategy more unambiguously fair, explaining the lower variation.
\update{}{Findings from prior studies on cognitive biases \S\ref{section:cog_bias} do not seem to apply significantly in our context (deciding how to split resources between groups with a 1:4 population ratio), as the vast majority of participants allocated resources fairly.}


Experiments 1 \& 2 allowed to examine the effect of framing on two presentation designs, one with text only, and one where text is supplemented with a visualization, with a congruent framing. In a follow-up experiment, we examine the role of each presentation component in isolation, as well as effects of congruent vs. \update{uncongruent}{incongruent} framings.


\section{Experiment 3: A Thorough Look at Information Framing}

The main purpose of the previous experiments was to investigate whether \emph{supplementing} text with a visualization influences fair allocations. In this experiment, we take a closer look at the role of information framing, exploring each presentation format independently (i.e text only, visualization only, and text plus visualization). This allows us to isolate the role of presenting information through visualization only. Prior studies found little influence of visualizations in inducing prosocial behaviors (\S \ref{sec:related_work}); in Experiment 3, we examine whether visualizations can substitute text, while promoting a similar propensity for fair allocations in our context. This also allows us to examine conditions where text and visualization have incongruent framings (e.g. text focuses on individuals and visualization focuses on group, and vice-versa, see \S \ref{sec:tool}). Experiment 1 \& 2 only included congruent framings (with additional caveat that text presentation included both group and individual framings in the individual-focused framing design).

\noindent Our main research question focused on the effect of visualization framing as follows\footnote{Note that our pre-registered material referred to `text' presentation format as `numerical', and `individual' framing as `person' framing.}:
\begin{enumerate}[itemsep=0pt, topsep=0pt, leftmargin=6em]
    \item[\update{}{\rint\ } \textbf{\small RQ3.1}] To what extent does visualization framing influence the unfairness of fund allocations across all text framing conditions? 
\end{enumerate}
\noindent Secondary research questions aimed at uncovering the effect of visualization framing were:
\begin{enumerate}[itemsep=0pt, topsep=0pt, leftmargin=6em]
    \item[\textbf{\small RQ3.2}] To what extent does visualization framing influence the fairness of allocations when
    \textbf{(a)} no text information is present;
    \textbf{(b)} text information is shown with a group framing;
    and when \textbf{(c)} text information is shown with an individual framing?
    \item[\textbf{\small RQ3.3}] Does the effect of visualization framing depend on text framing?  
\end{enumerate}
\noindent An additional secondary research question focused on examining the effect of presenting a visualization:
\begin{enumerate}[itemsep=0pt, topsep=0pt, leftmargin=6em]
    \item[\textbf{\small RQ3.4}] To what extent does the presence of visualization influence the fairness of allocations when
    \textbf{(a)} the text information and visualization have a congruent group framing;
    \textbf{(b)} the text information and visualization have a congruent individual framing;
    and when \textbf{(c)} the text information and visualization have an incongruent framing?
\end{enumerate}
Finally, our two auxiliary research questions were:
\begin{enumerate}[itemsep=0pt, topsep=0pt, leftmargin=6em]
    \item[\textbf{\small RQ3.5}] What are the most common stated reasons for the donation allocations?
    \item[\textbf{\small RQ3.6}] How does each condition affect the duration taken to complete an allocation choice?
\end{enumerate}

\subsection{Experimental Design}

We followed the procedure described in \S\ref{sec:procedure}. Program Alpha still had 3 children, but we changed the size of program Zefa to now include 120 children, resulting in a 1:40 ratio (instead of 1:4 in the other Experiments). Since a larger difference between the programs might be more prone to compassion fade, this will allow us to further examine the robustness of our previous findings. We also changed the amount of the resource to be allocated accordingly. Keeping a target of \$200 per individual for a fair allocation, the total amount to be allocated was changed from \$3,000 to \$24,600.

\begin{figure*}[t]
    \centering
    \includegraphics[width=\textwidth]{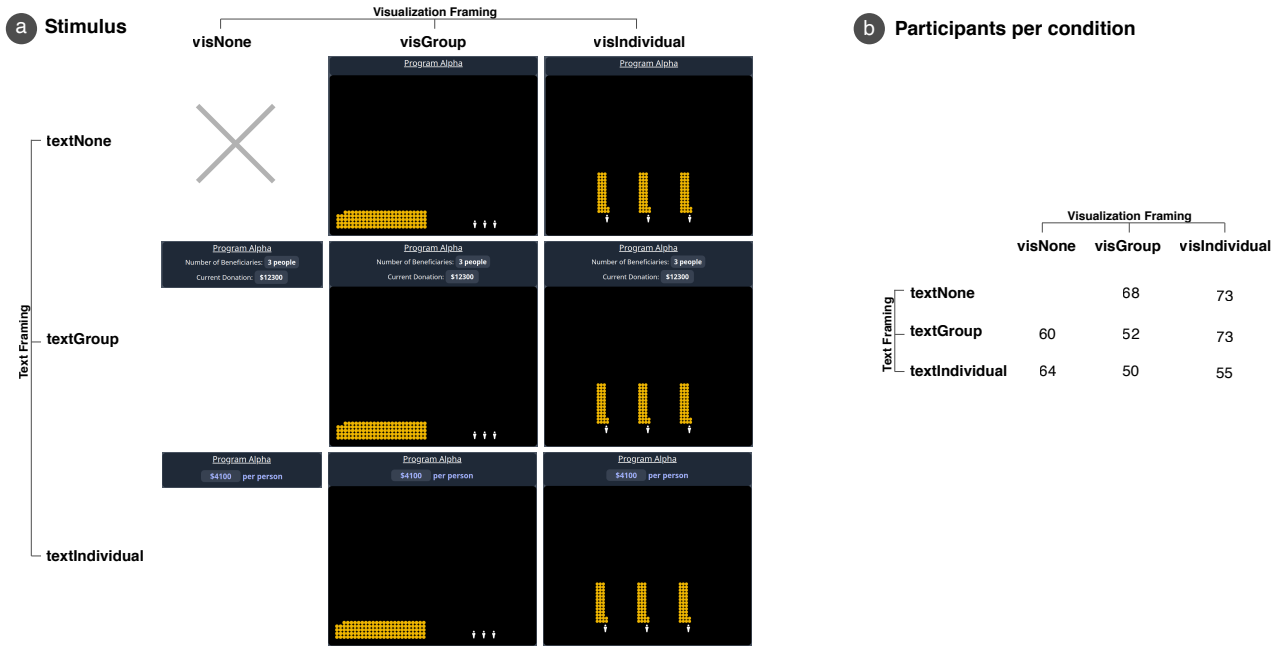}
    \caption[]{Experiment 3: (a) Stimulus used in the different conditions. See a demo of the tool at \url{https://resallocdesign.github.io/}. (b) Participant count per condition.}
    \label{fig:exp_3_conditions}
    \Description{Two figures (a) and (b). Figure (a) has nine images formatted in a 3x3 matrix. Each column shows images of conditions using different visualisation framing. The following conditions are shown in the order of left to right: visNone, visGroup, visIndividual. Each row shows images with different text framing. The following conditions are shown in the order of top to bottom: textNone, textGroup, textIndividual. The top left cell for condition has an "X" marked. The top left cell for condition has an “X” marked. All other images have two boxes. In the textNone row, each box has the program name label in text at the top. In the textGroup row each box has the program name, number of beneficiaries and current donation label in text at the top. For example: “program Alpha”, “number of beneficiaries: 3 people”, “current donation: \$12300”. In the textGroup row each box has the program name and amount allocated per person in text, for example "\$4100 per person". The first column on the left has no image underneath the text. The second column in the middle has boxes with a stack of yellow circles on one side and horizontally stacked human silhouettes on the other side, underneath the text. The third column on the right has boxes with a stack of yellow circles on top of the horizontally stacked human silhouettes, underneath the text. All circles represent 100 dollars in the text and each silhouette is 1 person. Figure (b) has the same matrix as (a) but only numbers in each cell. The numbers read, from top-left to bottom-right: blank, 68, 73, -next row- 60, 52, 73, -next row- 64, 50, 55.  }
\end{figure*}

\update{ We also made subtle refinements to the designs: w}{Our visual design was refined through unregistered piloting. W}e switched to a dark background to achieve higher contrast, and traded the coins pictogram for a simpler yellow circle for increased simplicity and readability (\autoref{fig:exp_3_conditions}).

We ran a \update{within}{between}-subject experiment were we manipulated \update{two independent variables.}{the presence and framing of both a text and a visualization presentation format as \update{independant}{independent} variables.}  The \textsc{Text-Framing} \update{}{variable} specifies the framing employed for the text presentation format: \textit{textGroup} uses a group-focused framing only; \textit{textIndividual} uses an individual-focused framing only; and \textit{textNone} means that text presentation is not included. \update{Similarly, the}{The} \textsc{Visualization-Framing} \update{}{variable} specifies framing of the visual representation of information, with \textit{visGroup} and \textit{visIndivudal} values corresponding to a group and an individual framing respectively, and \textit{visNone} corresponds to no visualization at all.

The full factorial design results in nine conditions. However, we discard the \emph{(textNone, visNone)} condition as it results in an empty display (i.e. no information is presented at all). The stimulus included in the display component of the interactive tools (\autoref{fig:interface}-b) for all conditions respectively are depicted in \autoref{fig:exp_3_conditions}-a.

\subsection{Participants}
Our target sample size was 512 participants ($\sim$ 64 per condition). \update{}{After a successful pilot with 6 participants which informed visual design refinments, we ran this study on June 26, 2022 until target sample size was reached.}
Seventeen (17) participants were excluded. 
The 495 valid responses were distributed across conditions as per \autoref{fig:exp_3_conditions}-b. 


\subsection{Measurement \& Analysis}

We performed the measurements and analyses as described in  \S\ref{sec:measurements} \& \S\ref{sec:analyses}. To gauge allocation (un)fairness, we define the \textit{unfairness} metric as the distance to the fair allocation, in dollars. A value of zero means a fair allocation has been made; the further from zero, the unfairer the allocation. For instance, an unfairness value of \$600 means that either \$24,600 or \$23,400 has been allocated to Zefa, both of which are a distance of \$600 away from a fair allocation.

We note that our unfairness metric is identical to the Gini index -- a standard measure of income inequality in populations ~\cite{gini1997concentration} --  up to a multiplicative constant.
We use this distance metric as we wanted a metric which preserves our exact allocation value for both programs as specific breakpoints might derive additional meaning in the context of allocation strategies.
\update{}{With this measure, we used statistical contrasts to generate the bootstrapped confidence intervals used to answer our research questions.}

\begin{figure*}[t]
    \centering
    \includegraphics[width=\textwidth]{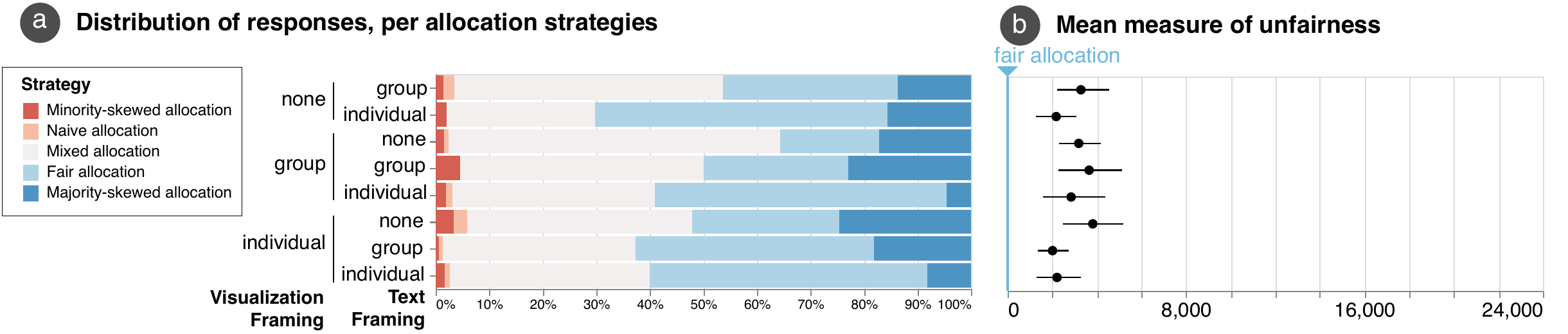}
    \caption[]
    {Results of Experiment 3: (a) Distribution of allocation strategy chosen by participants, by \textsc{Text Framing} and \textsc{Visualization Framing}, and (b) Mean measure of unfairness per condition. Error bars are 95\% CIs.}
    \label{fig:exp_3_summary}
    \Description{One stacked bar chart plot labelled (a) and one plot of eight confidence intervals with point estimates labelled (b). Plot (a) has the number of people using the five strategies mentioned in Experiment 1, for each of the eight conditions. For most conditions, mixed allocation is the majority, then fair allocation and then majority-skewed allocation. Plot (b) has confidence intervals for each of the eight conditions. Values range between 0 and 6000. Point estimates range between 2000 and 4000.}
    \vspace*{-1em}
\end{figure*}

\subsection{Results}
\autoref{fig:exp_3_summary} shows a summary of our results.  The stacked bar chart \texttt{(a)} shows responses per allocation strategy by condition. The plot \texttt{(b)} shows the mean unfairness measurement per condition. From \autoref{fig:exp_3_summary}-a, we note that fair allocations were not a majority in all conditions, suggesting potential effects in this more granular experimental design, which we investigate through our research questions as follows.

\begin{figure*}[t]
    \centering
    \includegraphics[width=\textwidth]{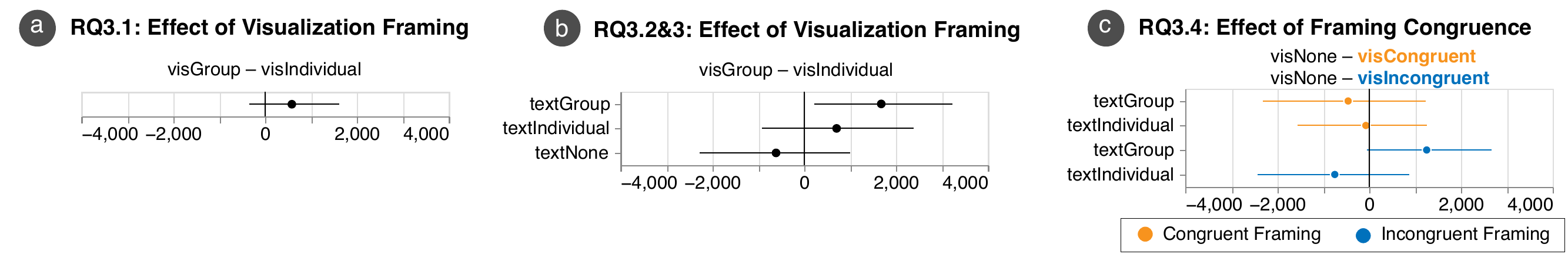}
    \caption[]
    {Results of Experiment 3: Mean difference between individual and group framing, overall (a), by text framing (b). Mean difference between no visualization and presence of visualization, by text framing and congruence of framings (c). Error bars are 95\% CIs.}
    \label{fig:exp_3_analysis}
    \Description{Figures labeled (a), (b), and (c). Figure (a) has a confidence interval and point estimates for the contrast between visGroup and visIndividual. The interval is between -1000 and 1500 with the mean near 500. Figure (b) has three intervals, the top interval is labeled textGroup is to the right of 0 and does not cross 0. The middle one is labeled textIndividual is to the right of 0 and does cross 0. Third one is the the left of 0 and crosses 0. Figure (c) shows four confidence intervals. The first two are intervals with a congruent text framing. Both intervals are to the left zero with a means close to zero. The third and fourth intervals for incongruent framing. The third interval is labeled textGroup and is to the right of zero but looks to touch or slightly cross zero. The fourth interval is label textIndividual and is to the left of zero and crosses it. }
    \vspace*{-1em}
\end{figure*}

\update{}{\rint} \textbf{RQ3.1: Effect of Visualization Framing.}
As our primary research question, we were interested in learning the extent to which visualization framing (\textit{visGroup} vs. \textit{visIndividual}) influences the unfairness of allocations across all text-framing conditions.
\autoref{fig:exp_3_analysis}-a shows the difference in mean allocation unfairness corresponding to the statistical contrast \textit{visGroup} $-$ \textit{visIndividual}, with \textit{visGroup} (resp. \textit{visIndividual}) corresponding to the union of all text framing conditions where visualization framing is group-focused (resp. individual-focused).
This represents the influence of visualization framing across all text framing conditions.
We found very weak evidence that group framing for visualizations might cause more unfairness in allocations when compared to an individual framing for visualizations.

\textbf{RQ3.2: Effect of Presenting Visualization Framing.}
We were interested in learning the extent to which visualization framing influences allocation fairness across all values in \textsc{Text-Framing}.
\autoref{fig:exp_3_analysis}-b shows the difference in unfairness between \textit{visGroup} and \textit{visIndividual} when presented with different text framings.

To find the effect of visualization framing when no text is present (RQ3.2a), we estimated the contrast \small{$\emph{(textNone, visGroup)}$} 
${\small - \emph{(textNone, visIndividual)}}$. To find the influence of visualization framing when text is shown with an \textit{individualText} framing (RQ3.2c), we estimated the contrast $\small{\emph{(textIndividual, visGroup)} - \emph{(textIndividual, visIndividual)}}$.
Since the CIs for both contrasts clearly cross zero, we find no evidence that group framing causes more or less unfairness in allocations when compared to individual framing for visualizations when no text, or when individual text framing is used.

To find the effect of visualization framing when text is shown with a \textit{groupText} framing (RQ3.2b), we computed the contrast $\small{\emph{(textGroup, visGroup)}} - \small{\emph{(textGroup, visIndividual)}}$.
Since the CI is to the right of zero, we find some evidence that group framing for visualizations causes more unfair allocations than individual framing for visualizations, when a group text framing is present.

\textbf{RQ3.3: Interaction Between Text Framing and Visualization Framing.}
We wanted to know the extent to which the effect of visualization framing depends on text framing.
To do this, we visually assess the overlap between the CIs in \autoref{fig:exp_3_analysis}-b.
The overlap between the CIs for \textit{textNone} and \textit{textIndividual}; and for \textit{textGroup} and \textit{textIndividual}, is too large to be able to conclude in an interaction \cite{krzywinski2013error}.
The overlap between the CIs for \textit{textGroup} and \textit{textNone} is relatively small, so there is some weak evidence that visualization framing has a larger effect (i.e, group visualization framing increases unfairness more as compared to individual visualization framing) when group text framing is used than when there is no text.

\textbf{RQ3.4: Effect of Presenting Congruent or Incongruent Visualizations.}
We were interested in learning the extent to which the presence of visualization influences allocation fairness when the framing between text and visualization was congruent and incongruent.
\autoref{fig:exp_3_analysis}-c shows the means difference in unfairness between the conditions where visualization framing is congruent and incongruent with text framing. In this chart, positive values can be interpreted as ``the presence of visualization results in more fair allocations than when there is no visualization'', in other words, adding a visualization helps; and negative values can be interpreted as ``the absence of visualization results in more fair allocations than when there is a visualization'', in other words, adding a visualization harms.

To find the effect of showing a visualization when the text and visualization have a congruent \emph{group} framing (RQ3.4a) we contrast $\small{\emph{(textGroup, visNone)} - \emph{(textGroup, visGroup)}}$ (top row in \autoref{fig:exp_3_analysis}-c). To find the effect of showing a visualization when the text and visualization have a congruent \emph{individual} framing (RQ3.4b) we contrast the conditions $\small{\emph{(textIndividual, visNone)} - \emph{(textIndividual, visIndividual)}}$ (\autoref{fig:exp_3_analysis}-c, second row). As both CIs for the difference of means clearly cross 0, we cannot claim the influence of congruent group or individual framings between visualization and text.

To find the effect of showing a visualization when text information and visualization have an incongruent framing (RQ3.4c) we preform two contrasts.
When \textit{textGroup} has an incongruent framing with the visualization, we contrast the conditions \\ $\small{\emph{(textGroup, visNone)} - \emph{(textGroup, visIndividual)}}$ (third row in \autoref{fig:exp_3_analysis}-c).
We find that the difference in means is largely within the positive range.
Even though the CI for the difference of means closely crosses zero, we can claim that when text presents allocations per group, adding a visualization presenting allocations per individual helps achieve fairer allocations.
When \textit{textIndividual} has an incongruent framing with the visualization, we contrast the conditions $\small{\emph{(textIndividual, visNone)} - \emph{(textIndividual, visGroup)}}$ (last row in \autoref{fig:exp_3_analysis}-c).
As the CI crosses zero, we find no evidence that individual framing for visualizations causes more or less unfair allocweakations than no visualizations, when text conveys information with a group framing.

\textbf{RQ3.5 \& RQ3.6: Auxiliary research questions.}
We were interested in the most common stated reasons for the donation allocations (RQ3.5). 
Since the number of participants using a mixed-allocation strategy differentiated significantly from past experiment, we decided on a further investigation of responses for this category only.
We performed an exploratory coding of the responses with 3 codes belonging to one category.
The \emph{zefa cause} code indicated whether a participant based their justification around giving more to the Zefa cause than to the Alpha cause; the \emph{alpha individual} code indicated the participant based their justification around giving more to the individuals in Alpha than to the individuals in Zefa; and \emph{balanced} code indicated the participant said they gave similar to both programs.  We found that most participants still justified giving more to Zefa due to the higher group size.

To find how each condition affects the duration taken to complete an allocation choice (RQ3.6) we plot the CIs for the mean allocation time for all conditions (see supplementary material). 
We found that most participants took between 35 and 85 seconds to reach their decision, and that participants in the \textit{(textNone, visIndvidual)} and \textit{(textIndividual, visNone)} conditions reached a decision the fastest.

\subsection{Discussion}

We took a closer look at the effect of framing. Overall, our findings suggest that individual framing, either for visualization or text, seems to provoke slightly fairer allocations when compared to group framing. This result reinforces the assumption that framing information, in a way that helps donors to identify the amount allocated to individuals, motivates them to split the resources equally among all the beneficiaries. 
We discuss specific effects in more detail.

We found that visualization framing can play a role on resource allocations in specific situations \rint. We found that including an individual-framed visualization is able to curb unfairness compared to group-framed visualizations, when the visual representation supplements text conveying allocations to groups. We suspect that the visualizations that displayed allocations per individual helped participants to identify the amount that each beneficiary would receive, making them to allocate the money more fairly than when presented with a group-framed text only. However, it is important to note that showing visualizations, either individual- or group-framed, without an accompanying text may have prompted participants to make more unfair allocations. Therefore, it seems that the presence of text influences participants to opt for a fair allocation strategy.

We also found that providing stand-alone text with an individual framing may produce similar fair allocations than including individual-framed visualizations combined with any text framing \rint. However, using group-framed text without including a visualization influences participants to have more unfair allocations. This finding is important since not only including a visualization with individual framing did not curb the effect of text framing (i.e. the visualization did not harm), it also helped in cases where text was presenting allocations with a group framing. Therefore, incorporating certain visualization framings may bring a net benefit in terms of promoting fairness, and at worse, not influence the donor. \update{}{This confirms that individual choices are prone to framing effects \cite{gosling2020interplay}, more specifically, that particular visualization framings have an influence on how people make decisions (\S\ref{section:vis}).} This finding adds contrast to prior work within our domain which has limited findings for visualizations which try to influence prosocial behaviours \cite{morais2021can}.

\section{Discussion and Future Work}

This section discusses our findings, limitations, and research implications.

\subsection{Summary of Findings}

%

This work explored the extent to which data visualization and information framing can promote fair resource allocation in humanitarian decision-making. 
We found evidence that allocations are more fair overall when resource allocation information is framed around individuals rather than groups \update{}{\rframe}. Our findings on the effects of adding visualizations are less conclusive \update{}{\rvis}, but suggest that visualizations may be more effective at promoting fairness when they are individual-framed and accompanied by a textual representation \update{}{\rint}. Those findings contribute to advance the understanding of framing in the context of information visualization and shed light on possible research directions. 

Our experiments are a first attempt to understand the interplay between visualizations and information framing in the context of an interactive decision-support tool for humanitarian resource allocation. We identified five different strategies people use to allocate resources (see \S\ref{sec:measurements}). Overall, we found that most participants chose \emph{fair} or \emph{mixed} allocations. 
Mixed allocations give slightly less weight to individuals in the larger group. Thus, the presence of a large number of mixed allocations suggests the presence of \update{a compassion fade effect \cite{butts2019helping}}{cognitive biases that may influence participants to favor the smaller population group (possibly exacerbated by a larger population ratio between the groups in Experiment 3)}. However, the even larger number of fair allocations suggests that in our particular experiment setting, compassion fade is relatively mild. Other cognitive biases such as the drop-in-the-bucket effect and the diversification bias do not seem to have played a major role in participants' decisions, since strategies  favoring individuals from the smaller group and \emph{naive} allocations giving equally to both programs were relatively uncommon.

\update{}{We find it interesting that, while individual framing achieved more fair allocation overall, the absence of strong bias which we could expect when emphasizing entatitivity somehow challenges our premise. } \update{}{
Cognitive biases may have had a more nuanced role in influencing allocation, however, this and prior work has not compared different sensitivities which participants may have for each bias. For example, some participants may have only been affected by \emph{drop-in-the-bucket} while others may have been only affected by \emph{compassion fade}. This work does not aim to prescribe specifically which biases visualizations are able to curb. As such, we recommend further investigation of the specific biases discussed in \S \ref{sec:related_work} in regards to resource allocation visualizations and framing.
}

Despite the relatively large proportion of fair allocations, unfair allocations were nonetheless common, and thus there is clearly room for improvement. It appears that the prevalence of fair allocations can be affected to some extent by how allocation information is presented. Our first and second experiment did not find conclusive evidence that adding a visualization to text using a congruent framing makes a clear difference. However, we found converging evidence that individual-framing causes a higher proportion of fair allocations, independently of the presentation format used. Thus it seems that individual-framing should be the choice if the goal is to promote fair allocations, while adding a visualization at least does not seem to harm. Thus, designers can choose to add a visualization to provide other benefits. We discuss those implications further in section \ref{subsec:what-does-this-mean}. 

In the last experiment, we investigated the effect of information framing in more detail, separating the effects by presentation format. In line with findings from our previous experiments, our results suggest that individual framing plays a clear role in promoting fair allocations. Also consistent with our previous experiments, we found that adding an individual-framed visualization to an individual-framed text (i.e., \emph{congruent individual framing}) does not seem to promote fairer allocations compared to individual-framed text alone. Similarly, we did not find evidence that adding a visualization with \emph{congruent group framing} yields less fair allocations. Thus, it does not seem that adding a visualization to a text reinforces the effect of information framing, if the framing is the same \update{}{which aligns with past work finding little benefit to visualization in eliciting prosocial behavior~\cite{morais2020showing, liem2020structure}}. However, we did find evidence that adding an individual-framed visualization to a group-framed text yields to more fair allocations overall. 



\subsection{Study Limitations}
\label{subsec:limitations}

Due to the complex and multi-faceted nature of humanitarian resource allocation problems, there were a number of trade-offs that we had to make to arrive to a tractable study.

First, our simplified task design is far from perfectly capturing real-world situations. In particular, our allocation scenario involves only two charity programs, where all targeted individuals have the same needs and whose life will be impacted equally if they get the same amount of financial aid. Those assumptions are unlikely to hold in virtually any real-word humanitarian resource allocation problem, if only because there are vast inequalities among individuals and because different humanitarian programs are generally far from being equally impactful~\cite{dragicevic2022information}. In addition, real-world humanitarian resource allocation problems involve complex interactions between multiple organizations and humanitarian programs, whereas our scenario assumed that the individuals from the two groups would not receive other forms of help. Similarly, personal charitable donations interact, and how fair a specific donation is depends on what other people have donated and to whom. It is important to note that our tool is only a stylized tool whose goal is to understand how people make decisions in controlled settings, and that a real decision-making tool should take into account the many additional dimensions of humanitarian resource allocation.


\update{}{Our work suggests that interactive visualization tools may hinder some cognitive biases in the context of humanitarian decision-making, but data visualizations can also mislead and provoke additional biases~\cite{szafir2018good}. The stages of the visualization pipeline can induce perceptual or cognitive misconceptions~\cite{mcnutt2020surfacing}: curating flawed data, making wrong data transformations, using deceptive visual encodings, misreading the visualization, etc. Our studies investigated how visualization design and information framing can contribute to curb cognitive biases related to decision-making, but other biases that affect judgement on perceptual or social levels were not directly addressed~\cite{calero2018studying}. More work is needed to investigate the impact of visualization tools in charitable decision-making and biases ingrained in this process.}

It is important to note that in our study, we used a simplified notion of fairness, as fairness is a complex concept. We chose to scope and narrow the notion of fairness as providing equal financial aid to all individuals to make our study tractable; and we developed metrics around that definition. We note that our proposed metric designed for Experiment 3 is equivalent to the standard Gini index; and was therefore a reasonable choice. However, more elaborate metrics that account for the many factors that real resource allocation problems encompass and for the diversity of normative ethical frameworks will be needed to study allocation fairness more thoroughly. Future work should use more holistic measures that capture the multi-dimensionality of allocation fairness, and generalize to an arbitrary number of groups.

\subsection{Perspectives for Visualization}
\label{subsec:what-does-this-mean}

This paper brings together theories of framing, cognitive biases, and visualization to explore the effect of design choices for interactive humanitarian resource allocation tools. It is still premature to prescribe visualization design principles or guidelines for reducing unfairness in resource allocations. However, our findings advance our understanding of the problem, potential solutions, opportunities, and challenges in this area.

To start with, our findings add nuance to previous studies which found little influence of visualizations in inducing behaviors that benefit others \cite{morais2021can}. We found that visualizations do not harm in our context, and can help in some circumstances.
This is an important result as beyond decision making tasks, visualizations have a wealth of potentially beneficial properties -- they draw attention~\cite{hohman2020communicating}, they are easier and faster to consume and interpret, and they can even promote self-reflection~\cite{chi1989self}, making them an important asset when communicating about resource allocations.

Although our work only presents preliminary findings about how to design visualizations for resource allocation tools, we have evidence that some design choices may facilitate good resource allocation decisions. Broadly, visualizations should be designed to emphasize help or impact on each individual instead of (or in addition to) on a global scale. It is however important to note that the visualization designs we tested are only one of the many possible designs that could be imagined. In particular, future visualization designs could be imagined that convey richer information. 
For instance, the pictorial marks representing group populations in anthroprographics \cite{morais2020showing} could be made visually richer so as to capture the diversity of population samples instead of impersonal standard iconography. Other resources such as time or the actual impact on individuals could be represented instead of coins. 

When designing visualizations for future resource allocation tools, one challenging question will be what additional information to include, and how. For instance, would representing geographical locations, genders, races, nature and severity of the problem, be important to convey? Or could it instead be counter-productive to add certain information, due to human biases? The design space for mitigating cognitive bias in visualization~\cite{wall2019toward} helps to critically reflect on such questions, and on possible design strategies to consider, such as temporarily hiding certain information as not all data is always helpful. We see the exploration of these questions as an exciting avenue for future research.

\subsection{Perspectives for HCI}
\label{sec:where-do-we-go}

Our resource allocation user interface has been designed and deployed as an experimental tool to investigate the role of different design dimensions on resource allocation decision-making. As such, it integrates only very basic functions. In general, our study can be seen as an initial exploration of the vast design space of user interfaces for such tools. There are many opportunities that we see our work could be extended to further support decision-makers.

First, we did not push interactive design beyond implementing a slider to dynamically control the allocation. The HCI literature can provide inspiration to extend our approach to foster particular behaviors, such as by inserting visual nudges~\cite{matejka2016effect}, adding stickiness~\cite{wall2019toward}, or repeatedly nudging the donor~\cite{mota2020desiderata, cockburn2020framing}. Further, creative, engaging interaction paradigms could activate curiosity and playfulness in tools dealing with otherwise serious problems. There might be potential for such an interface to intrigue and incite people to learn more about causes, their differences, their needs, and how they, as individuals, can help in an effective manner. \update{}{A systematic review of the tools that are currently employed in large associations and governments managing multiple charitable programs would also allow to better understand current practices, and challenges that decision makers face with the current tools.}

Second, resource allocation tools in the future could benefit from some level of automation. For instance, scraping content online to populate the interface with real-world, timely data, or to plug them into databases such as {\small usaspending.gov}.
Future tools could also encompass computational models, for instance, by incorporating predictions which could enable donors to appreciate how their donation can make concrete change. 

We also see an opportunity to elevate such resource allocation tools from tool to partners~\cite{grudin2017tool}, to achieve collaborative intelligence between the decision-maker and the computational tool~\cite{wilson2018collaborative}. A first step would be to look at recent work in improving automation and heuristics-based assistance for the allocation of resources for humanitarian aid \cite{aiken2022machine, anparasan2019resource, callaway2022rational}. Further, the utility of resource allocation tools is not simply for donors themselves but could also be used for visualizing an automated amount given out by some system, such as one integrating a machine learning model. Such tools provide ways to alleviate potential harms in algorithmically-infused societies \cite{wagner2021measuring}. Again, we are excited by the many possibilities to extend our work.
\section{Conclusion}

This work investigated how different aspects of tool design, namely presentation format and information framing, can nudge people into allocating humanitarian resources fairly. Our results indicate that there may be benefits in using text and visualizations with certain framings for such resource allocation tools. We found that, in general, individual-framed resource allocation information (i.e., showing how much each individual will receive) seems to be more effective in promoting fair allocations, whether information is conveyed through text or through a visualization. We also found that adding an individual-framed visualization to a group-framed text (i.e., that shows how much each group will receive) promotes fair allocations. However, showing individual-framed text appears to be sufficient, and fairness does not seem to be improved by adding individual-framed visualizations.
While most of our results are tentative, they are a first step to better understand the interplay between information visualization and framing for the purposes of humanitarian resource allocation.

For the sake of experimental control and in order to make our study tractable, we made a number of simplifying assumptions. We used a fictional resource allocation scenario with two charity programs that are equivalent except for the number of individuals they they target, and we assumed that all individuals have equal needs and will not receive help from other organizations. Future work should explore resource allocation problems that are more realistic, for example by involving charity programs that differ across several dimensions and populations that are more diverse. In addition, we only focused on the resource allocation problem for two groups. In real-world examples, charities oftentimes need to allocate resources to more groups. Therefore, further studies should also generalize our methods to three or more groups. Finally, none of the designs we tested was able to eliminate occurrences of mixed allocations, and therefore it seems likely that some degree of compassion fade persists across designs. 
While our study looked at a very specific scenario, we hope it will serve as a starting point for a wider range of research work.



\bibliographystyle{abbrv-doi}

\bibliography{bibliography}

\end{document}